\documentclass[prd,superscriptaddress,onecolumn,showpacs,11pt,msmath,preprintnumbers,showkeys]{revtex4}
\usepackage{wrapfig,rotating}
\usepackage{graphicx,epsfig,color}



\usepackage{graphicx}
\usepackage{dcolumn}
\usepackage{bm}

\def\beq{\begin{equation}}
\def\eeq{\end{equation}}
\def\beeq{\begin{eqnarray}}
\def\eeeq{\end{eqnarray}}

\def\2GPD{$_2\mbox{GPD}$}

\def\12{$1\otimes 2$}
\def\22{$2 \otimes 2$}
\def\eff{{\mbox{\scriptsize eff}}}

\def\Qsep{Q_{\mbox{\rm\scriptsize sep}}}
\def\Qsep2{Q^2_{\mbox{\rm\scriptsize sep}}}

\begin{document}
 \title{Heavy Quark Energy Loss in the Quark-Gluon Plasma in the Moller theory.}
 \author{B.\ Blok$^{1}$,\\[2mm] \normalsize $^1$ Department of Physics, Technion -- Israel Institute of Technology,
 Haifa, Israel\\}
 \begin{abstract}
\par We study the energy loss of a heavy quark propagating in the Quark-Gluon Plasma (QGP) in the framework of the Moller 
theory, including possible large Coulomb logarithms as a perturbation to BDMPSZ  bremsstrahlung, described in the Harmonic Oscillator (HO) approximation.
 We derive the analytical expression that describes the energy loss in the entire emitted gluon frequency region.
In the small frequencies region, for angles larger than the dead cone angle, the energy loss is controlled by the BDMPSZ mechanism,
while for larger frequencies it is described by N=1 term in the GLV opacity expansion. We estimate corresponding quenching rates
for different values of the heavy quark path and different $m/E$ ratios.

 \end{abstract}

   \maketitle
 \thispagestyle{empty}

 \vfill

\section{Introduction.}
\par The energy loss of heavy quarks propagating through the media was widely discussed in recent years in different formalisms.
In particular the heavy quark energy losses were intensively studied in the 
BDMPSZ \cite{BDMPS1,BDMPS2,BDMPS3,Z0A,Z0B,Z0C} approach,  starting from 
\cite{DK}.
\par The authors of \cite{DK} assumed that like in the vacuum, the heavy quark radiation is suppressed by the dead cone effect,
\beq
\omega\frac{ dI^{\rm vac}}{d\omega dk^2_t}\sim \frac{\alpha_sC_F}{\pi^2}\frac{k^2_t}{(k^2_t+\theta^2\omega^2)^2},
\label{dc}
\eeq 
where $I$ is the multiplicity of the heavy quark, $\theta=m/E$ is the dead cone angle, m is the heavy quark mass,   $\omega$ and $\vec k_t$  are  the frequency and the transverse momenta of the radiated gluon, and $E$ is the energy of the heavy quark.
 They resulting  heavy quark quenching rate is then significantly smaller than the one observed in the experiment, where up to rather small energies of order  several masses of heavy quark,
 the jet quenching rates of the heavy and light/massless quarks are the same \cite{wang,kurt}.
 \par However it was found in \cite{ASW,Z}  ( see also \cite{ADSW,DG,ZHQ,BT} for related research) that the dead cone effect is actually absent, both in the Harmonic Oscillator (HO) approximation  to the BDMPSZ approach and in the first N=1 term in the  GLV  opacity expansion \cite{GLV,GLV1,Wiedemann}.
 This observation leads to significant increase of  heavy quark jet quenching rate, however there still exists the overall decrease the jet quenching factors  with mass, and the problem is not solved.
  In fact the simulations carried in \cite{ASW} and \cite{Z} show that 
 in both HO and GLV approximations the quenching rate is approximately  constant up to $\theta\sim 0.05$ and then starts to fall.
  \par In another interesting development it was pointed in \cite{arnold,arnold1,Z1,Z}  that the interference pattern in the parton propagation is  determined by the minimal of the two available coherence lengths -the LPM coherence length, the quantum diffusion formation length 
  and the  parton path length $L$.
  Let us define the  quenching coefficient as 
  \beq
  \hat q =\int \frac{d^2q_t}{(2\pi)^2}nq^2\frac{d\sigma^R_{el}}{d^2q_t},
  \eeq
  where $n$ is the density of the scattering centres in the media, and $\sigma^R$ is the scattering cross section of the projectile parton in the color representation R.
  The transverse momenta accumulated  in the diffusion regime is 
  \beq
  k^2_t\sim \hat q l_{c},
  \eeq
  where $l_{c}$ is a diffusion coherence length, corresponding to Landau Pomeranchuk Migdal (LPM) effect,
 $l_c^{LPM}(\omega )=\sqrt{\omega/\hat q}$.  Next there is the quantum diffusion formation length, similar to heavy quark propagating in the media
  without interference between different media scattering centres, $l^q_c\sim 1/(\theta^2\omega)$.   The actual physical radiation regime is determined
  by the shortest of these lengths \cite{Z,arnold1}.
  
  \par   For light quark, as it was shown in \cite{arnold1} for frequencies much smaller than $\omega_c\sim \hat qL^2$, the dynamics of the quark is determined by LPM interference. This interference is usually described in the so called Harmonic Oscillator (HO) approximation. On the other hand for frequencies $\omega\ge \omega_c$, the energy loss is described by the N=1 GLV formalism \cite{GLV,GLV1,Wiedemann}. The reason why, although there may be a lot of elastic scatterings, the use of a first term in the opacity expansion  is still justified is rather straightforward \cite{arnold1}: the N=1 GLV approximation corresponds to the tail of $q^2_t$ probability distribution, i.e to the regime when the large   but rare momentum transfers  are dominant.
   
   \par For heavy quark  for the small frequencies the gluons are emitted  outside the dead cone and can be described in the same HO approximation as for light quarks. However for frequencies larger
   than $\omega_{DC}=(\hat q/\theta^4)^{1/3}$, when the gluons start to be emitted inside the dead cone region, the quantum diffusion lenth $l_c^q=1/\theta^2\omega$  starts to be smaller than the diffusion coherence length  $l_{coh}^{LPM}$ \cite{Z1} and  the dynamics of radiation for these large frequencies is determined by N=1 GLV approximation.

   \par In a further development the authors of \cite{mehtar,mehtar1}  obtained the formula for light quarks, that explicitly describes not only the  diffusion and N=1 GLV regime, but also 
   the intermediate region of frequencies, and  thus  is applicable to the dynamics of light quark quenching in the entire frequency region. Their
   formula takes into  account possible  Coulomb interaction corrections to the LPM bremsstrahlung, treated as the perturbation.
   \par In the current paper we shall generalise the results of \cite{mehtar,mehtar1} to heavy quarks and obtain the unified formula that describes   the gluon radiation for arbitrary  frequencies. 
   The basic approach will be to build the perturbation theory for heavy quark around the HO approximation. We shall see that while the radiation beyond dead cone is determined, for  
   $\omega\le \omega_{DC}$ by LPM effect and is similar to the one for light quarks, while the dead cone radiation is described by N=1 GLV approximation.
   \par Throughout the whole paper we shall assume that the dominant gluons are soft, $\omega=xE,x<<1$. It is quite simple to include the finite frequencies using $\omega=z(1-z)E$, however the
   full calculation will then require also the inclusion of the space phase constraints, that will make the calculations much more complicated.
   \par The paper is organised  in the following way. In the section 2 we describe the dynamics of heavy quark propagation in the media, in section 3 we review the description of heavy quark in the HO
  In section 4 we build a perturbation theory for heavy quarks and derive the explicit expression for the energy loss. 
 We use this expression to estimate qualitatively the heavy quark energy loss in section 5, and to estimate the quenching weights in section 6. 
  Our results are summarised in section 7.

 \section{Heavy Quark Propagation in the QGP}
 \subsection{Basic formalism}
 The heavy quark energy loss  in the media is given by
 \beq
 \omega \frac{dI}{d\omega }=\frac{C_F\alpha_s}{(\omega)^2}2Re\int ^\infty_0dt_1\int^{t_1}_0 dt \partial_{\vec x}\partial _{\vec y}(K(\vec x,t_1,\vec y,t)-K_0(\vec x,t_1;\vec y,t))\vert_{\vec x=\vec y=0}.
\label{e1}
 \eeq
 Here $K$ is the propagator of the  particle in the media with the two dimensional effective potential due. to the scattering  centres, and 
 $K_0$ is the corresponding propagator of the free particle in the vacuum. The effective two dimensional potential is given by 
 \beq
 V(\vec\rho )=i\int \frac{d^2q_t}{(2\pi)^2}(1-\exp(i\vec q_t\vec\rho))\frac{d^2\sigma_{el}}{d^2q_t}.
 \eeq 
 Here $d^2\sigma_{el}/d^2q_t$ is the cross section of elastic scattering of high energy particle 
 on the media centre. The  media is described by Gyulassy-Wang model\cite{GW}.
 The effective potential in the momentum space is given by 
  \beq
\frac{ d\sigma(\vec q_t )}{d^2 q_t}=\frac{4\pi\alpha_sm^2_DT}{(q_t^2+\mu^2)^2}\equiv \frac{g^4n}{(q_t^2+\mu^2)^2},
 \eeq
 where  the parameter $\mu\sim m_D$, and the Debye mass $m_D$ is given by 
 \beq
 m_D\sim  4\pi\alpha_sT^2(1+N_f/6)=\frac{3}{2}g^2T^6
 \eeq
 for $N_f=3$ light quarks, T is the media/QGP temperature. The density of the scattering centres in the GW model is given by $n=\frac{3}{2}T^3$,
 and  the strong coupling is $\alpha_s=\frac{g^2}{4\pi}$.
 The effective potential in the coordinate space is 
  \beq
 V(\rho)=\frac{\hat q}{4N_c}(1-\mu \rho K_1(\mu \rho )=\frac{\hat q\rho^2}{4N_c}(\log(\frac{4}{\mu^2\rho^2})+1-2\gamma_E)\label{pot},
 \eeq 
 where $\gamma_E=0.577$ is the Euler constant,  and the bare quenching coefficient is
 \beq
 \hat q=4\pi\alpha_s^2N_cn.
 \eeq
 \subsection{Perturbation Theory}
 \par  For processes that are dominated by large momentum transfer oit is enough to
 take into account only the first terms in the Taylor expansion of $V(\rho)$. 
 The first approximation  corresponds to the quadratic term in the expansion \ref{pot} and is called the HO (harmonic oscillator ) approximation.
 In this approximation the effective potential V is given by
 \beq
 V(\rho)=\frac{1}{4}\hat q_{\rm eff}\rho^2\label{lpm}.
 \eeq
Here $\hat q_{\rm eff}$ is the  effective jet quenching coefficient, given by 
\beq
\hat q_{\rm eff}=\hat q\log( \frac{Q^2}{ \mu^2})\label{eff},
\eeq
and Q is the typical transverse momenta, accumulated by the particle on the scale of the coherence length.

\par The HO effectively describes the LPM bremsstrahlung \cite{BDMPS1}. More 
precise treatment of the energy loss includes also large Coulomb logarithms and is called in the theory 
of the Abelian (QED) LPM effect the Moller theory \cite{katkov}. In the QCD framework the inclusion of Coulombic interactions 
can be made using the perturbation theory \cite{mehtar,mehtar1}. Namely, instead of the usual opacity expansion
\cite{GLV1,GLV,Wiedemann}, we shall consider the perturbation theory around the oscillator potential adding the Coulombic effects as a perturbation.
The effective potential in Moller theory is given by 
\beq
V(\rho) =\frac{1}{4}\hat q\rho^2\log(1/\rho^2\mu^2),
\eeq
and includes the short range coulombic  logarithms. In the framework of the perturbation theory this potential is split as 
\beq
V(\rho)=V_{HO}(\rho)+V_{pert}(\rho), V_{HO}(\rho)=\frac{\hat q\log(Q^2/\mu^2)}{4}\rho^2,V_{\rm pert}(\rho)=\frac{\hat q}{4}\log(\frac{1}{Q^2\rho^2}),\label{pert}
\eeq
where Q is the typical momenta, defined above, equal to $Q\sim \sqrt{\hat q\omega}$ in the HO approximation.
We shall need sufficiently large Q, so that 
\beq
\log(Q^2/\mu^2)\gg \log(\frac{1}{Q^2\rho^2}),
\eeq
i.e. perturbation theory is applicable meaning that we probe rather small transverse distances.
\par Then the energy loss is given by  Eq. \ref{e1},
where the propagator K is calculated in perturbation theory as  \cite{mehtar,mehtar1}
\beq
K(\vec x,t_1;\vec y,t)= K_{HO}(\vec x,t_1;\vec y,t)-\int d^2z\int^{t_1}_t dsK_{HO}(\vec x,t_1;\vec z,s) V_{pert}(z)K_{HO}(\vec z,s;\vec y,t_1)\label{15}
\eeq
Here $K_{HO}$ is the heavy  quark propagator in the imaginary two dimensional potential $V_{HO}$:

\begin{eqnarray}
K_{HO}(\vec x,t_1;\vec y,t)&=&\frac{i\omega\Omega}{2\pi \sinh \Omega (t_1-t)}\exp(\frac{i\omega 
\Omega }{2}\{\coth\Omega  (t_1-t)(\vec x^2+\vec y^2)-\nonumber\\[10pt]
&-&\frac{2\vec x\vec y }{\sinh\Omega (t_1-t)}\})\exp(-i\theta^2\omega (t_1-t)/2),\nonumber\\[10pt]
\label{e2}
\end{eqnarray}
and 
\beq
\Omega=\frac{(1+i)}{2}\sqrt{\hat q/\omega}
\eeq
In the limit  when there is no media this propagator reduces to free quark propagator 
\beq
K_0(\vec x,t_1;\vec y,t)=\frac{i\omega}{2\pi }\exp(i\frac{\omega (\vec x-\vec y)^2}{2(t_1-t)}).\label{e3}
\eeq
\subsection{Qualitative Dynamics of the Heavy Quark}
\par The expansion written in the form \ref{15} clearly exhibits the formation lengths described in the Introduction:
the heavy quark mass leads to the oscillating exponent $\exp(i\theta^2\omega/2 (t_1-t))$ in Eq. \ref{e2}, while the harmonic oscillator part of the propagator \ref{e2} oscillates 
with the frequency $\sqrt{\omega/\hat q}$. Then it  is clear that when $l_c^q<< l_c^{LPM}$ the oscillations due to heavy quark mass cut off the integral for heavy quark
energy loss, the  oscillating harmonic oscillator part of the propagator is approximately freezed and the LPM effect is not relevant, the energy
loss is defined by the induced radiation on the scattering centres-the N=1 GLV. On the other hand, in the opposite case, the heavy quark exponent
is close to one, and the integral for energy loss is controlled by the HO multiplier. We have LPM bremsstrahlung plus corrections due to coulomb logarithms.
\par We can now choose the substruction scale $Q$ in the momentum space. As it was explained in \cite{arnold1,mehtar}  this scale corresponds to the typical momentum accumulated 
by the quark along the coherence length propagation. Such momentum squared  is $\hat q\times\sqrt{\omega/\hat q}$ for $\omega<<\omega_{DC}$ and $\sim \theta^2\omega^2\sim \omega/l^q_c$
for $\omega>>\omega_{DC}$. Consequently we shall use the interpolation formula
\beq 
Q^2=\sqrt{\omega \hat q_{\rm eff}}U(-\omega+\omega_{DC})+\theta^2\omega^2U(\omega -\omega_{DC}),\label{m1}
\eeq
where $U(x)$ is a unit step function:$U(x)=1$ if $x\ge 0$, and $U(x)=0$ if $x\le 0$.
We shall use another interpolation formula to check the sensitivity to the exact $Q$ value in the intermediate region around $\omega_{DC}$:
\beq
Q^2=\sqrt{\hat q_{\rm ef}\omega+\theta^4\omega^4}.\label{m2}
\eeq
\par Alternatively, the dynamics of the heavy quark can be approached using the arguments in \cite{arnold1}. Namely , in the LPM (diffusion ) regime the distribution
over momentum transfers in the scattering on the media centres is described by a gaussian, peaked in the $Q^2_{typ}\sim \sqrt{\hat q w}$. The scattering with significantly higher 
momentum transfers $q_t$ is described by the tail of the distribution, which is N=1 GLV, that essentially describes the independent coulomb scattering on the media centres.
In this region the LPM gaussian is parametrically close to zero, and N=1 GLV dominates. It was explained in \cite{Z1} that N=1 term in opacity expansion is a good description of large momentun transfer regime, since such
scatterings in the tail occur quite rarely. Since inside dead cone the typical momenta is  $k^2_t\sim \omega/l_c^q\sim \theta^2\omega^2\gg \sqrt{\hat q\omega}$, inside the dead cone we shall find ourselves in 
the GLV regime.
\subsection{N=1 GLV}
\par We shall also need the explicit expression for N=1 term in the opacity expansion for massive quark. The corresponding result was 
derived in \cite{ASW}, and has the form:
\begin{eqnarray}
\omega \frac{dI}{d\omega} &=&\int dk_t^2\int_0^\infty dq^2\frac{4\alpha_sC_F\hat q q k_t}{\pi\omega}\frac{ LQ_1-\sin(LQ_1)}{Q_1^2}\frac{q^2}{q^2+\theta^2\omega^2}\nonumber\\[10pt]
&\times&\frac{m^2_D(k^2+\theta^2\omega^2)+(k^2-\theta^2\omega^2)(k^2-q^2)}{(k^2+\theta^2\omega^2)((m^2+k^2+q^2)^2-4k^2q^2)^{3/2}}.\nonumber\\[10pt]
\label{GLV}
\end{eqnarray}
where 
\beq
Q_1=(q^2+\theta^2\omega^2)/(2*\omega).
\eeq
Here $k_t$ is the momentum of the radiated gluon.

\section{Heavy Quark in the HO Approximation}

\par Let us review the leading order contribution to the energy loss of heavy quark, that in our perturbation approach corresponds to
Harmonic approximation. There are two parts in the expression \ref{e1} due to different regions of integration in $t_1$, we shall call them the 
bulk and the boundary contributions since in one case the integration in $t_1$ goes from 0 to L and in the second from $L$ to  $\infty$.
Note that the authors of \cite{mehtar,mehtar1} used different approach due to results in \cite{arnold} that permits for massless case the calculation of the integral 
\ref{e1} without splitting into two regions. However, it is not clear how to extend the method of \cite{arnold} to the  case.of massive quarks.
We shall review here the heavy quark energy loss calculation in HO approximation and represent the results in the form of the one dimensional integrals.

\par \subsection{HO Bulk Contribution.}
This term is equal to
\beq
\omega\frac{dI^{\rm HO\,\,\,Bulk}}{d\omega}=\frac{\alpha_s}{\omega^2} 2Re\int^L_0dt_1\int^{t_1}_0 dt\partial_{\vec x}\partial_{\vec y}(K(\vec x,t_1;\vec y,t)-K_0(\vec x,t_1;\vec y,t))\vert_{\vec x=\vec y=0},
\eeq
where $K_0$ is the propagator of the free heavy quark, and $K_{HO}$ is the heavy quark  propagator in the HO approximation given by  Eq. \ref{e2}.

After differentiation we obtain in the soft gluon limit:
\beq 
\omega\frac{d^{\rm HO\,\,\,Bulk}}{d\omega}=-\frac{\alpha_s C_F}{\pi}  2Re\int^L_0 dt_1\int ^{t_1}_0 dt (\frac{\Omega^2}{(\sinh( \Omega (t_1-t))^2}-\frac{1}{(t_1-t)^2})\exp(-i\theta^2\omega((t_1-t)/2) \label{eq1}
\eeq
Note that the integrand  is a function of $\tau=t_1-t$. We use the identity
\beq
\int_0^Ldt_1\int^{t_1}_0dsf(s)=\int^L_0(L-s)f(s)ds \label{identity2}
\eeq
to go from the double to one-dimensional integrals.
This means
\begin{eqnarray}
\omega\frac{dI^{\rm HO \,\,Bulk}}{d\omega}&=&\frac{-2\alpha_sC_F}{\pi}2Re\int ^L_0dt_1\int _0^{t_1}d\tau(\frac{\Omega^2}{\sinh(\Omega \tau)^2}-\frac{1}{\tau^2})Exp(-i\tau\theta^2\omega/2)\nonumber\\[10pt]
&=&\frac{-2\alpha_sC_F}{\pi}\int ^L_0ds (L-s)(\Omega^2/\sinh(\Omega s)^2-1/s^2)\exp(-is\theta^2\omega/2)\nonumber\\[10pt]
\label{b12}
\end{eqnarray}
\par In the limit  of the massless quark  $\theta->0$ we get the spectrum 
\beq
\omega\frac{dI}{d\omega}=\frac{2\alpha_sC_F}{\pi}Re\log(\frac{\sinh(\Omega L)}{\Omega L})\label{lb1}
\eeq
in agreement with  the BDMPSZ results for the bulk part of the spectrum for massless quark.

\subsection{HO Boundary term}

\par It is also easy to calculate the boundary term in the HO approximation:
\beq
\omega\frac{dI^{\rm HO\,\,Boundary}}{d\omega}=\frac{\alpha_s}{\omega^2} 
2Re\int^\infty_L dt_1\int^{L}_0 dt\partial_{\vec x}\partial_{\vec y}(K(\vec x,t;\vec y,t_1)-K_0(\vec x,t_1;\vec y,t)\vert_{\vec x=\vec y=0}\label{hob}
\eeq
The propagator K  in Eq. \ref{hob} corresponds to the new regime  when the particle travels outside of the media, $t_1>L,L>t>0$ . Consequently  it
 is given by by the convolution
\beq
K(\vec x,t_1;\vec y,t)=\int d^2z K_0(\vec x,t_1;\vec z,L)K_{HO}(\vec z,L;\vec y,t)
\eeq
Using the explicit expressions for $K_0$ and $K_{HO}$ given by Eqs. \ref{e2},\ref{e3} we obtain
\begin{eqnarray}
\partial_{\vec x}\partial_{\vec y }K(\vec x,t_1;\vec y,t)\vert_{\vec x=0,\vec y =0}&=&\int d^2z\int ^\infty_Ldt_1\int ^L_0dt \frac{1}{(2\pi)^2}\frac{\omega^4\Omega^2z^2}{(t_1-L)^2\sinh(\Omega(L-t))^2}\nonumber\\[10pt]
&\times&\exp(iz^2(\frac{\omega}{2(t_1-L)}
+\frac{\omega\Omega}{2} \coth\Omega(L-t)))exp(-i\theta^2\omega(t_1-t)/2)\nonumber\\[10pt]
\label{bt1}
\end{eqnarray}
It is easy to carry the integration over $d^2z$.
We have
\beq
\int d^2z\exp(iAz^2)z^2=\frac{\pi}{A^2}
\eeq
So we obtain
\begin{eqnarray}
\omega \frac{dI^{\rm HO\,\,\,Boundary}}{d\omega }&=&\int_L^{\infty} dt_1\int ^L_0dt (2Re\frac{1}{\pi}\frac{\Omega^2}{(t_1-L)^2\sinh(\Omega(L-t)^2}\frac{\exp(-i\theta^2\omega(t_1-L+L-t)/2)}{(1/(t_1-L)+\Omega \coth \Omega (L-t))^2}\nonumber\\[10pt]
&-&2Re\frac{1}{\pi}\frac{\exp(-i\theta^2\omega(t_1-L+L-t)/2)}{(t_1-t)^2})\nonumber\\[10pt]
\label{hob1}
\end{eqnarray}
we now can define  $s=t_1-L$, and take integral over s

 using  the formula 
\beq
\int^{\infty}_0ds\exp(-iAs)/(1+Bs)^2=\frac{B-iA\exp(A/B)*\Gamma (0,iA/B))}{B^2}
\eeq
where $\Gamma(s,x)$ is the incomplete gamma function  \cite{AS}.
We  obtain 
\begin{eqnarray}
\omega \frac{dI^{\rm HO \,\,\,Boundary}}{d\omega }&=&2Re\frac{\alpha_sC_F}{\pi}\int^L_0ds((\frac{2\Omega }{\sinh(2\Omega s)}-\frac{i\theta^2\omega}{2} \exp(i\frac{\theta^2\omega\tanh(\Omega s)}{2\Omega}))\nonumber\\[10pt]
&\times&\frac{\Gamma(0,\frac{i\theta^2\omega\tanh(\Omega s)}{2\Omega})}{\cosh(\Omega s)^2}\nonumber\\[10pt]
&-&(\frac{1}{s}-\frac{i\theta^2\omega}{2}\exp(i\theta^2\omega s/2)\Gamma(0,i\theta^2\omega s/2))\nonumber\\[10pt]
\label{b1}
\end{eqnarray}
In Eq. \ref{b1} 
 \beq
\Gamma(0,x)=-Ei(-x)-i\pi
\eeq
and  $s=L-t$, the function   Ei is the Integral Exponent function \cite{AS}.
Note that.  the  integrand in \ref{b1} is concentrated near the end of the media region, i.e. near $t\sim L$.
\par For small frequencies outside the dead cone the energy spectrum almost does not change when we take into account the quark mass, while for large frequencies in the 
dead angle region the spectrum decreases rather rapidly, in agreement with Dokshitzer-Kharzeev results.
\par The full HO result for massive quarks is then given by the sum of Eqs. \ref{b1} and \ref{b12}:

\beq
\omega\frac{dI^{\rm HO}}{d\omega}=\omega\frac{dI^{\rm HO\,\,\, Bulk}}{d\omega}+\omega\frac{dI^{\rm HO\,\,\, Boundary}}{d\omega}\label{cor31}
\eeq
\par For light quark  (i.e. in the $\theta\rightarrow 0$ limit) we have
for the bulk term
\beq
\omega \frac{dI}{d\omega}=\frac{\alpha_sC_F}{\pi}Re\log(\frac{\sinh (\Omega L)}{\Omega L})
\eeq
and 
\beq
\omega \frac {dI}{d\omega}=\frac{\alpha_sC_F}{\pi}(Re\log(\vert \cosh(\Omega L)\vert)-Re\log(\frac{\sinh (\Omega L)}{\Omega L})
\eeq
for the boundary term.
In the sum we obtain the famous BDMPS spectrum
\beq
\omega \frac{dI}{d \omega}=\frac{2\alpha_sC_F}{\pi}(\log(\vert\cosh(\Omega L)\vert)
\eeq
confirming the self consistency of our approach.

\section{Coulombic correction.}
We are now in position to calculate the corrections due to Coulomb logarithms.
As in the previous section we split the integration in Eq. \ref{e1} into two parts $0<t_1<L$  (the bulk term)  and $t_1>L$-the boundary term.
\subsection{Bulk Term}
\par 
. We start from the bulk term.
The Coulombic correction to the propagator for the heavy quark is given by  \cite{mehtar}
\beq
K_{pert}(\vec x,t_1;\vec y,t)=--\int d^2z\int^L_0 dt_1\int ^{t_1}_0 dt \int^{t_1}_t ds K_{HO}(\vec x,t_1;\vec z,s)V_{\rm pert}(\vec z,s)K_{HO}(\vec z,s,;\vec y,t)
\eeq
where  the perturbation potential is taken as in Eq. \ref{pert}
\beq V_{\rm pert}=\frac{\hat q}{4}\log{1/(z^2Q^2)} \eeq
where Q is the substraction point in momentum space, that must be taken as the typical momentum acquired in the set of elastic scatterings over the coherence length scale .  Note that the potential is not dependent on $s$.
As a result we have after differentiating the propagator over its endpoints,
\begin{eqnarray}
\omega \frac{dI^{\rm Bulk\,\,Coulomb}}{d\omega}&=&\frac{\alpha_sC_F}{\omega^2}2Re\int^L_0dt_1\int^{t_1}_0 dt \int d^2z\int_t^{t_1} ds  \nonumber\\[10pt]
&\times&\frac{\hat q}{4}\frac{\omega^4\Omega^4\exp(-i\theta^2\omega(t_1-t)/2)}{(2\pi)^2\sinh\Omega(t_1-s)^2\sinh\Omega(s-t)^2} \nonumber\\[10pt]
&\times&\exp(\frac{i\omega\Omega}{2} z^2(\coth(\Omega(t_1-s)+\coth\Omega(s-t)))z^4\log(\frac{1}{z^2Q^2})\nonumber\\[10pt]
\label{basic1}
\end{eqnarray}
We now carry the integration over $d^2z$ using
\beq
\int d^2z z^4\log(\frac{1}{z^2Q^2})\exp(-Bz^2)=
\frac{(-3+2\gamma_E+2\log(\frac{B}{Q^2}))\pi}{B^3}, Re(B)\ge 0.\label{int1}
\eeq
I
We then obtain
\begin{eqnarray}
\omega \frac{dI^{\rm Bulk\,\,\,Coulomb}}{d\omega}&=&-\frac{4\hat q\alpha_sC_F\Omega}{\pi\omega}2Re\int^L_0 dt_1\int^{t_1}_0 dt \int_t^{t_1} ds  \sinh\Omega (t_1-s)\sinh\Omega(s-t) \nonumber\\[10pt]
&\times&(i(-3+2\gamma +2\log(\frac{\omega\Omega}{2Q^2}\frac{\sinh\Omega(t_1-t)}{\sinh\Omega(t_1-s)\sinh\Omega. (s-t)]}))+\pi)\nonumber\\[10pt]
&\times&\frac{\exp(-i \theta^2\omega(t_1-t)/2}{\sinh(\Omega(t_1-t))^3}.\nonumber\\[10pt]
\label{10}
\end{eqnarray}
The  integral over $s$ can be. taken analytically in the limits between $t_1,t$. Quite remarkably under regularisation when we integrate between $t+\epsilon,t_1-\epsilon$,
and then take the limit $\epsilon\rightarrow 0$
the integral is finite. We then use 
the identity  \ref{identity2} , since the integrand depends only on the difference  $t_1-t$, to represent the Coulomb correction to the bulk
 contribution in the form of an one dimensional integral

\begin{eqnarray}
\omega\frac{dI^{\rm Bulk\,\,\,Coulomb}}{d\omega}&=&Re\int ^L_0dx (L-x)   2i\frac{\alpha_s C_F\hat q}{\omega\pi}\frac{\exp(i\theta^2\omega (-i\theta*\omega x/2)}{\sinh\Omega (x)^3}\nonumber\\[10pt]
&\times&\cosh(\Omega x)(-((-2+A+Log(4)-2\log(\Omega*\omega/Q^2))\nonumber\\[10pt]
&+&2*(\Omega x+\log(1-Exp(-2\Omega x)))-\log(2)))\tanh[\Omega x)\nonumber\\[10pt]
&-&(-\pi^2/6-(2+A)\Omega x-2Li_2(2,\exp(-2\Omega x))-Li_2(2,1)-(-i\pi+2\Omega x)^2/2\nonumber\\[10pt]
&-&\pi^2/3+2\Omega x(-i*\pi+\log((1-\exp(-2\Omega x))+2\Omega x+\log(-\Omega \omega/Q^2)),\nonumber\\[10pt]
\label{cor101}
\end{eqnarray}
where 
\beq
A=-i\pi+3-2\gamma_E.
\eeq
\par Here $Li_2$ is the dilogarithm (Spence) function \cite{AS}:
\beq
Li_2(z)=-\int^z_0\frac{\log(1-u)}{u}.
\eeq
Note that , since $\Omega$ is complex, the integrand is the complicated analytical function of its arguments. We have checked that this function has no discontinuities related  to the cuts of logarithm and dilogarithm
in the complex plane in the region of integration and as a function of $\omega$ and $\hat q$.

\subsection{The Correction to the Boundary Term.}

Next  we need to calculate the correction to the HO boundary  term. This correction  is given by the integral
\begin{eqnarray}
\omega\frac{dI^{\rm  Coulomb\,\,\, Boundary}}{d\omega}&=&\frac{\alpha_sC_F}{(2\pi)^3\omega^2}\frac{q}{4}\int^{\infty}_L dt_1\int_0^L dt \int^L_t ds \int d^2z\int d^2u\nonumber\\[10pt]
&\times&\frac{\exp(i\frac{\omega z^2}{t_1-L})}{(t_1-L)^2} (\vec z\vec u)\nonumber\\[10pt]
&\times&\exp(i
\frac{\omega\Omega (z^2*\coth(L-s)+u^2*\coth(L-s)-\frac{2\vec z\vec u}{\sinh(L-s)})}{2})\nonumber\\[10pt]
&\times&\frac{\exp(iu^2\coth(s-t)\frac{\omega\Omega}{2})}{\sinh\Omega(L-s)(\sinh \Omega(s-t))^2}\nonumber\\[10pt]
&\times&\omega^5\Omega^3 u^2\log(\frac{1}{Q^2u^2}) .  \nonumber\\[10pt]
\label{cor6}
\end{eqnarray}
The integral $\int d^2z$ is gaussian and can be easily taken, the remaining integral over $\int d^2u$ is taken using 
Eq. \ref{int1}.
\par After the integrations over $d^2z$ and $d^2u$ the resulting integral has the form
\begin{eqnarray}
\omega\frac{dI^{\rm  Coulomb\,\,\, Boundary}}{d\omega}&=&\int^\infty_L dt_1\int ^L_0dt\int^L_t ds\frac{i\alpha_sC_F\hat q\Omega}{\pi\omega} \exp (i\theta^2\omega (t-t_1)/2)\nonumber\\[10pt]
&\times&(A-2\log(\frac{\omega\Omega(\cosh(\Omega(L-t))\Omega(t_1-L)+\sinh(\Omega(L-t))}{2Q^2\sinh(\Omega(s-t))(\Omega(t_1-L)\cosh(\Omega(L-s))+\sinh(\Omega(L-s)))}))\nonumber\\[10pt]
&\times&\sinh(\Omega(s-t))(\Omega(t_1-L)\cosh(\Omega(L-s))+\sinh(\Omega(L-s))),\nonumber\\[10pt]
\label{cor41}
\end{eqnarray}

where 
\beq
A=-i\pi+3-2\gamma_E.
\eeq
The integral over s can be taken using Mathematica, since the integrand is essentially the rational function.
We then obtain the double integral o ver $t,t_1$:
\begin{eqnarray}
\omega\frac{dI^{\rm  Coulomb\,\,\, Boundary}}{d\omega}&=&\frac{i\alpha_sC_F\hat q}{2\pi\omega} \exp i\theta^2\omega (t-t_1)\nonumber\\[10pt]
&\times&\cosh(\Omega(L-t)) (2\log(((1-\exp(-2\Omega(L-t)))+\Omega(t_1-L)\nonumber\\[10pt]
&\times&(1+\exp(-2\Omega(L-t))))/(2\Omega (t_1-L)))\nonumber\\[10pt]
&+&2\Omega(L-t))\Omega (L-t_1)-((-2+A+\log(4)-2\log(\frac{\Omega\omega}{Q^2}))+2(\Omega (L-t)\nonumber\\[10pt]
&+&\log((1-\exp(-2\Omega(L-t))))-\log(2))))\tanh(\Omega(L-t))\nonumber\\[10pt]
&+&(-\pi^2/6-(2+A)\Omega (L-t)-Li_2(\frac{\exp(-2\Omega (L-t))(1+\Omega(L-t_1))}{1-\Omega (L-t)}\nonumber\\[10pt]
&-&Li_2(\exp(-2\Omega (L-t))-Li_2(-1+\frac{2}{1+\Omega (L-t_1))})-(-i\pi+2\Omega(L-t))^2/2\nonumber\\[10pt]
&-&(-i\pi+2\Omega (L-t)+\log(\frac{1+\Omega(t_1-L)}{1+\Omega(L-t_1)})^2/2-\pi^2/3\nonumber\\[10pt]
&+&2\Omega (L-t)(\log((\exp(-2\Omega (L-t)+\frac{-1+\Omega (L-t_1)}{1+\Omega (L-t_1)})+2\Omega (L-t)+\log(-\frac{\Omega\omega}{Q^2})))\nonumber\\[10pt]
&\times&(-1+\Omega (L-t_1)\tanh (\Omega (L-t))))/(\Omega (t_1-L)\cosh(\Omega(L-t))+\sinh\Omega(L-t))^3.\nonumber\\[10pt]
\label{cor8}
\end{eqnarray}

The final answer for the coulombic correction is then given by a sum of Eqs. \ref{cor101},\ref{cor8}:
\beq
\omega\frac{dI^{\rm Coulomb}}{d\omega}=\omega\frac{dI^{\rm Bulk\,\,\,Coulomb}}{d\omega}+\omega\frac{dI^{\rm Coulomb\,\,\,Boundary}}{d\omega}.
\label{cor30}
\eeq
We shall also need $\omega\frac{dI^{\rm Coulomb \,\,\, reduced}}{d\omega}$ which is given by Eq. \ref{cor30} but without the common factor $\hat q$.
The final answer  for the energy loss is the sum of Eqs. \ref{cor30},\ref{cor31}
\beq
\omega\frac{dI}{d\omega}(\omega,L,\hat q,Q)=\omega\frac{dI^{\rm HO}}{d\omega}(\omega,L,\hat q_{\eff},Q)+\hat q \omega\frac{dI^{\rm Coulomb\,\,\, reduced}}{d\omega}(\omega,L,\hat q_{eff},Q).
\label{cor15}
\eeq
where $\hat q_{eff}$ is given by Eq. \ref{eff} and the typical momenta $Q$ is given by Eqs. \ref{m1},\ref{m2}. Equation \ref{cor15} is our main result.

\par The integrals  \ref{cor101},\ref{cor8} are rather complicated. However we checked numerically that in the limit of massless quarks $\theta\rightarrow 0$ our results coincide with the ones
obtained in \cite{mehtar,mehtar1}. We were not able to reduce the expressions above to the light quark case analytically. However we checked that the zero mass quark expression derived in \cite{mehtar}
\beq
\omega \frac{dI^{\rm Coulomb}}{d\omega}=\int^L_0 ds \frac{1}{k(s)}\log(k(s)+\gamma),\label{cor12}
\eeq
where
\beq
k(s)=\frac{i\omega \Omega}{2}(\coth(\Omega(s)+\tanh(\Omega (L-s))
\label{12}
\eeq
coincides with \ref{cor15} numerically for all possible values of $L,\omega$
\par . We have also checked that  the expression \ref{12} can be easily 
derived also summing bulk and boundary contributions, instead of using the approach of \cite{arnold}.
\section{Numerical results}

\begin{figure}[t]
\includegraphics[scale=0.53]{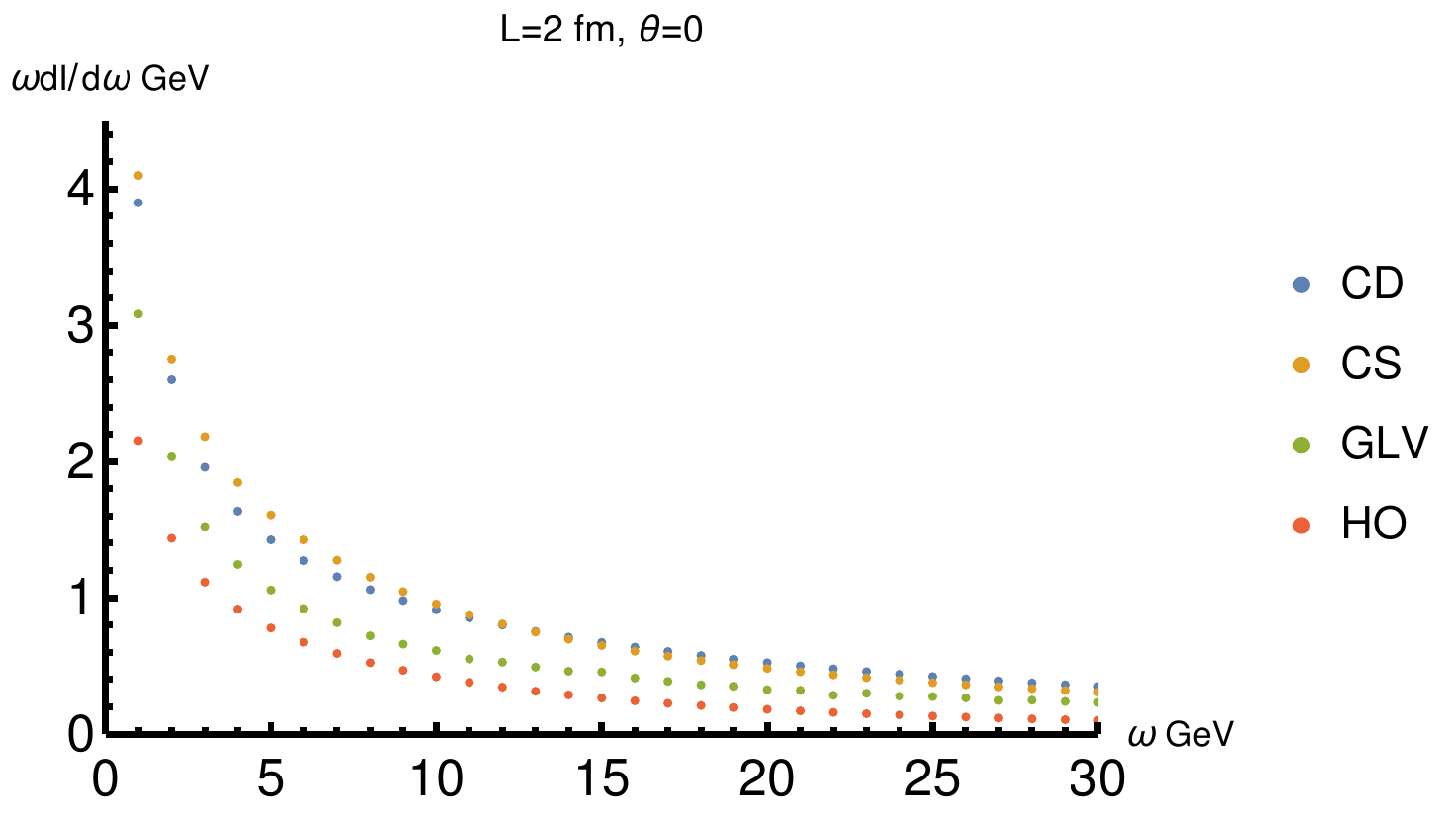}
\includegraphics[scale=0.53]{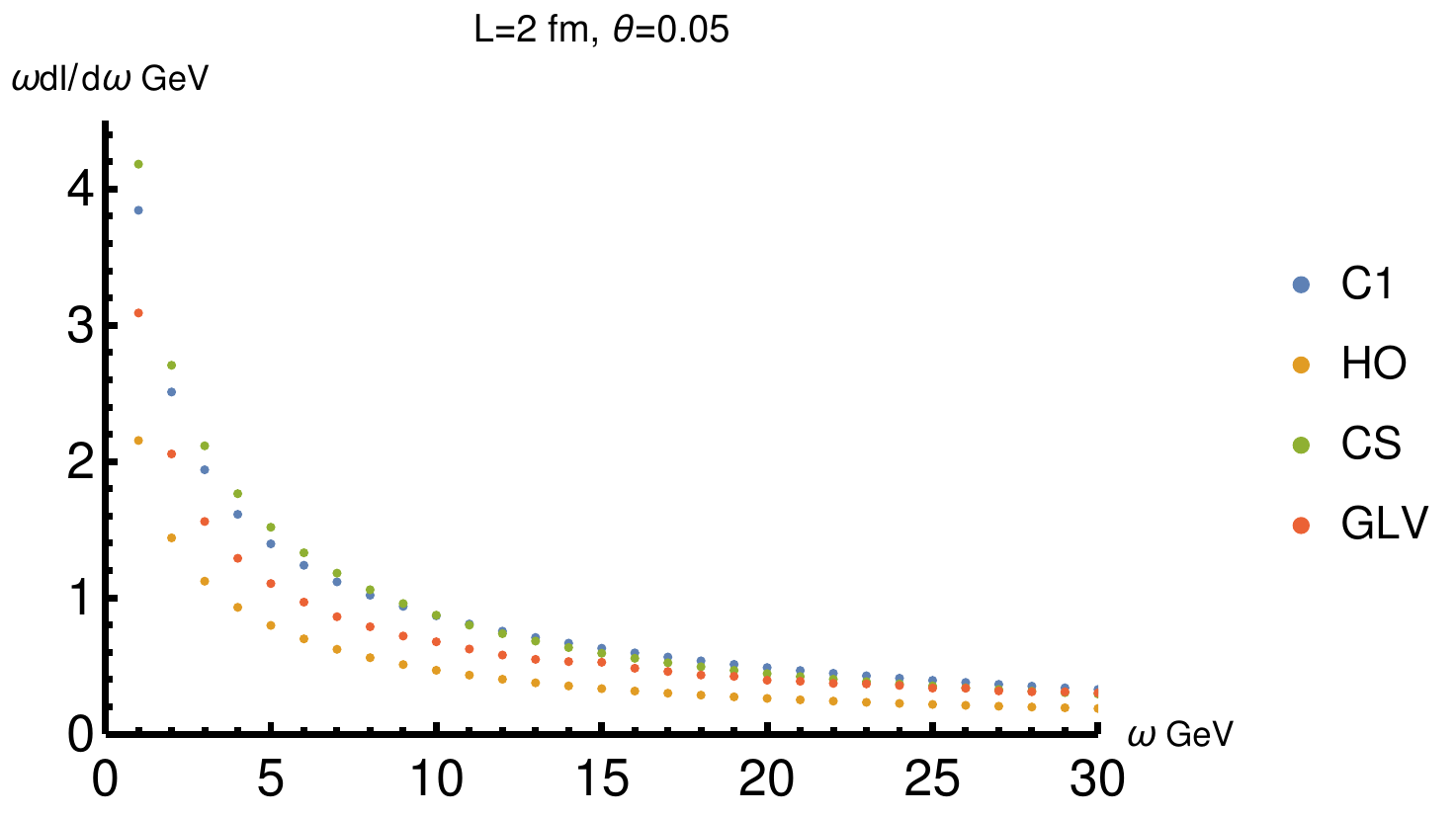}
\includegraphics[scale=0.53]{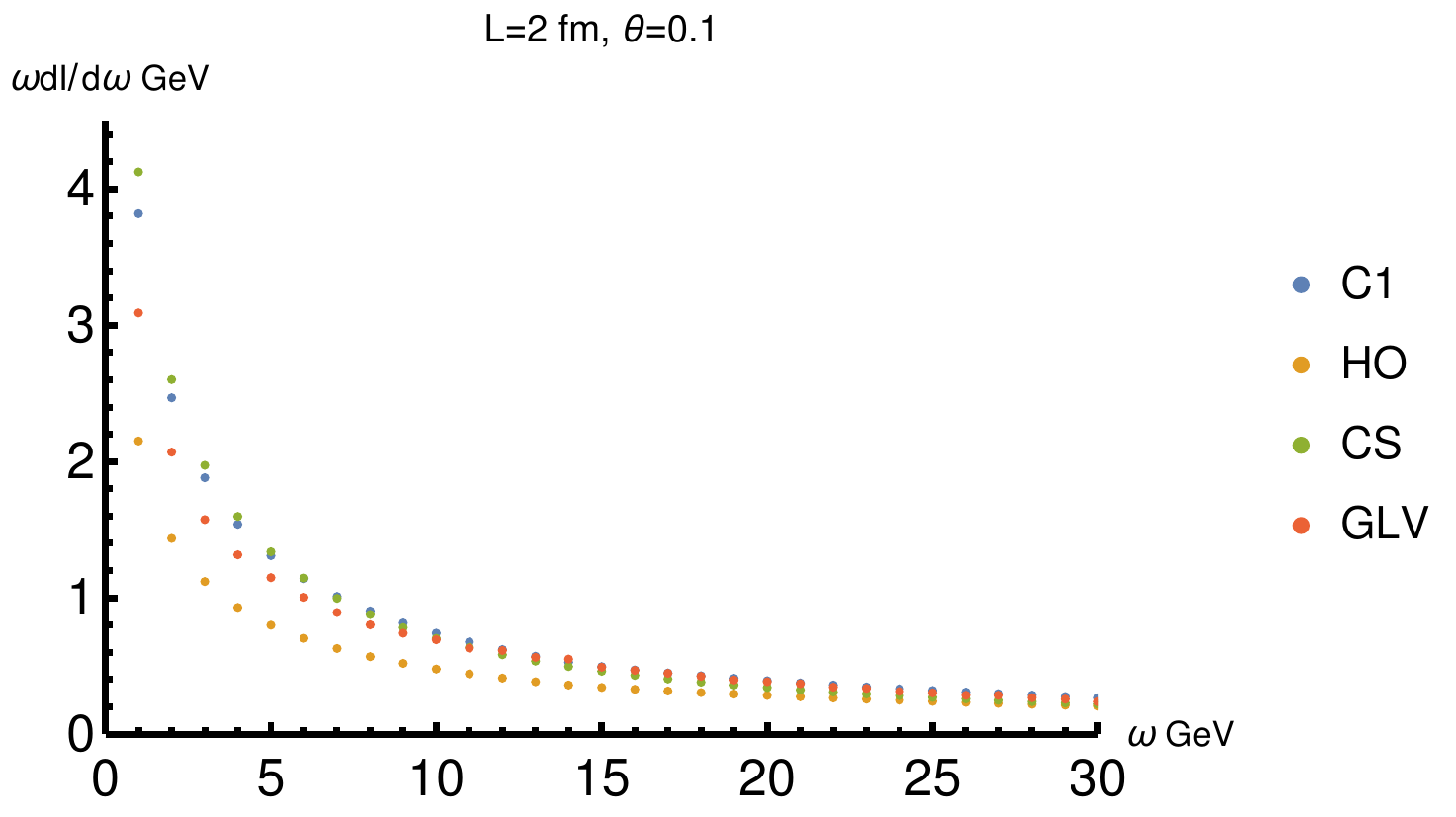}
\includegraphics[scale=0.53]{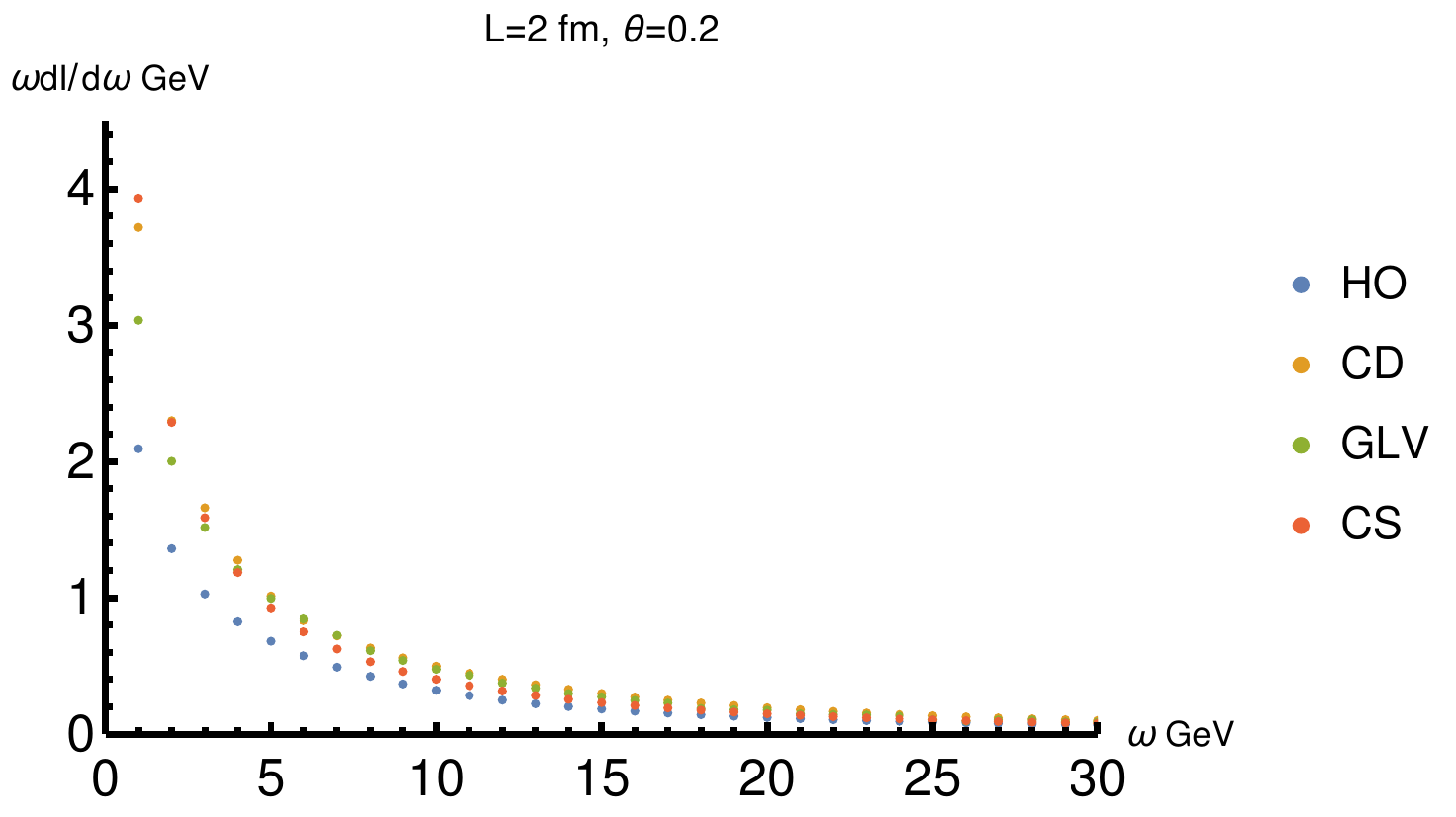}
\caption{\textsl{. The energy loss  in the leading order in $\alpha_s$ for L=2 fm for different values of $\theta$ as a function of the radiated gluon energy $\omega$, divided by $\alpha_sC_F$ for different
parametrisations of  $Q(\hat q,\omega,\theta)$  for HO and total (HO+Coulomb) contributions: CD, CS  are the total (HO+Coulomb)  energy losses  for typical momenta $Q$ given by Eqs. \ref{m1},\ref{m2} respectively;HO refers to HO approximation 
with $Q$ given by Eq. \ref{m1a}, and GLV refers to N=1 GLV expression with $Q$ independent $\hat q$.}}
\label{1}
\end{figure}

\begin{figure}[t]
\includegraphics[scale=0.53]{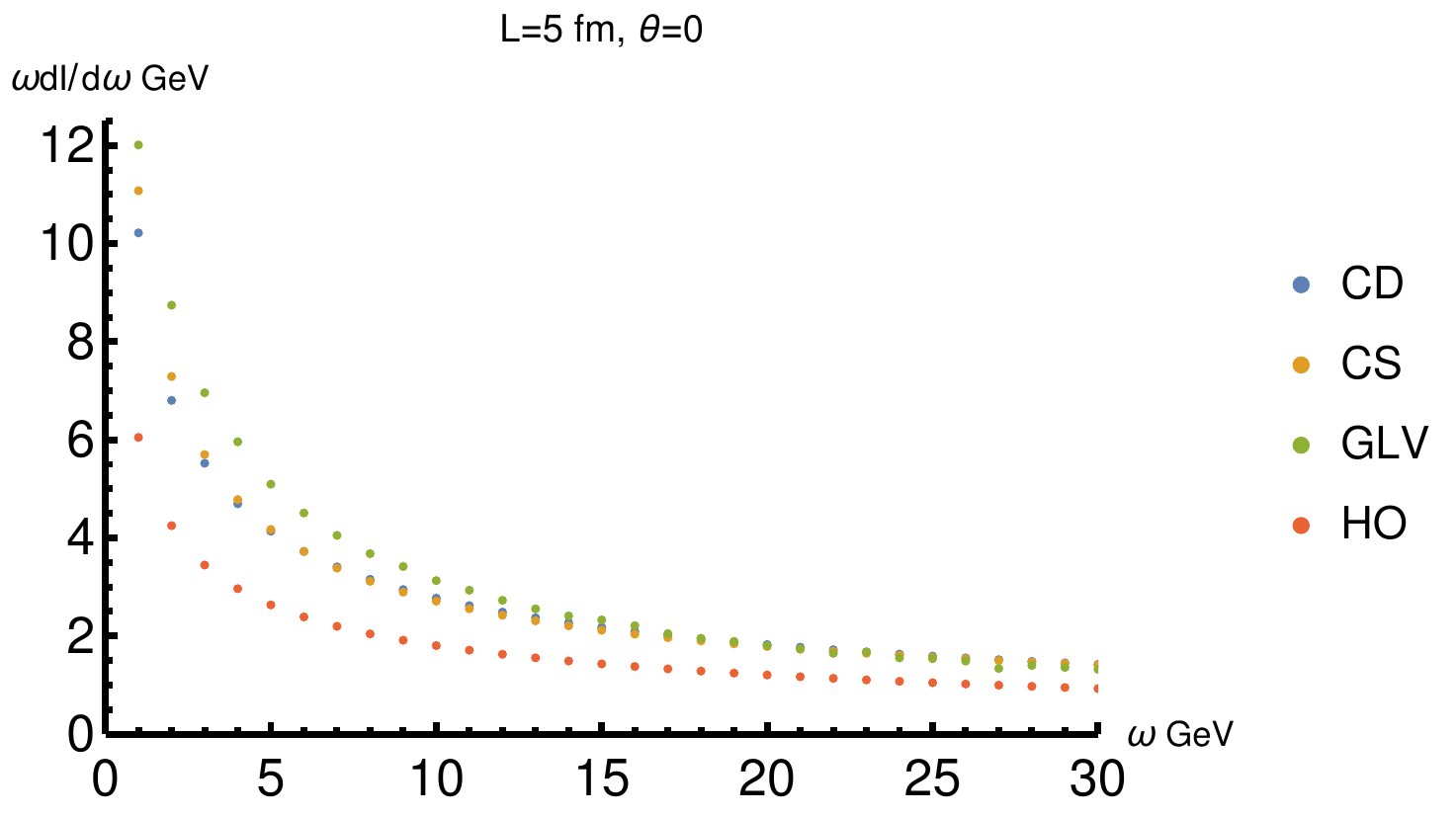}
\includegraphics[scale=0.53]{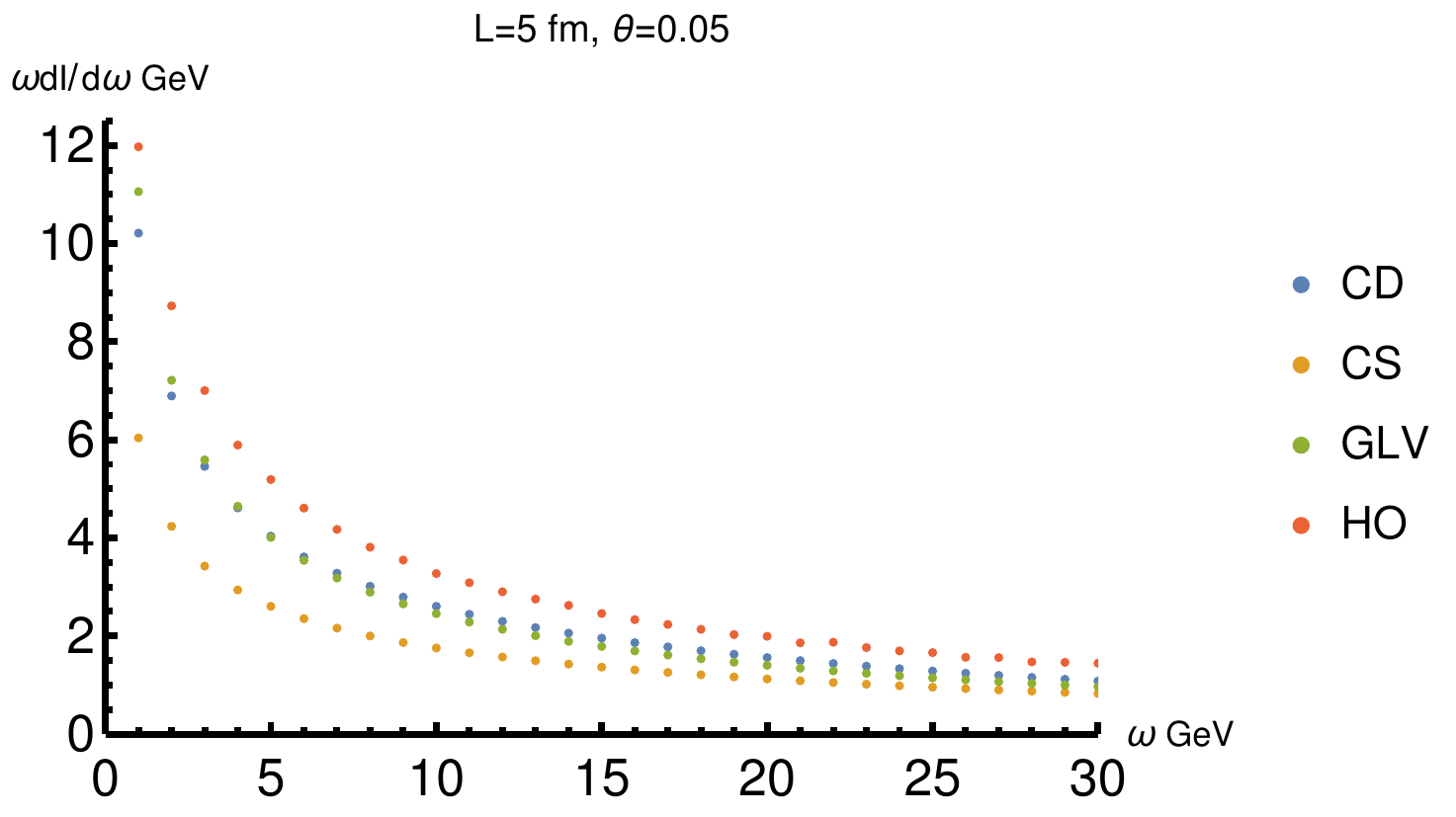}
\includegraphics[scale=0.53]{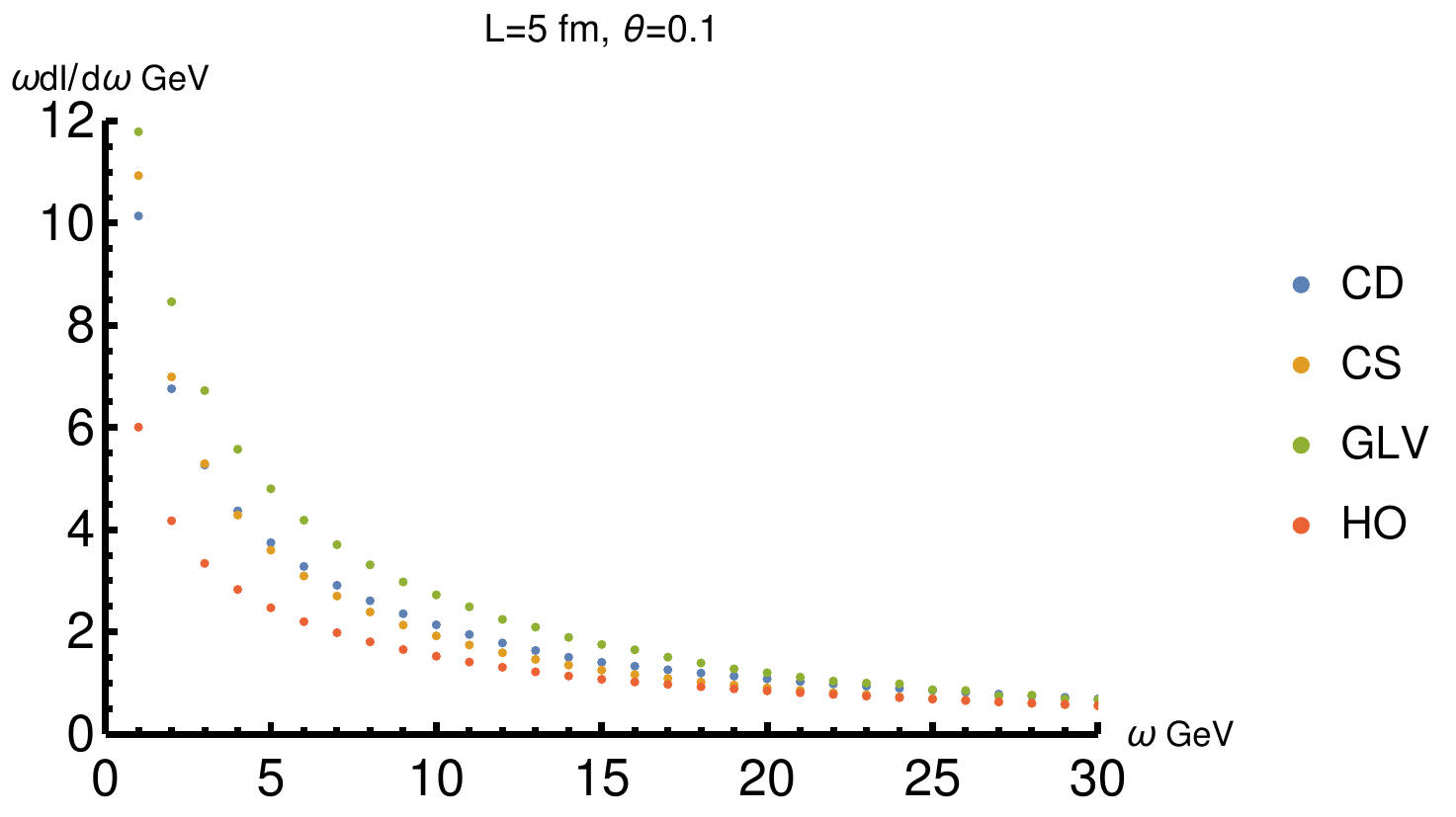}
\includegraphics[scale=0.53]{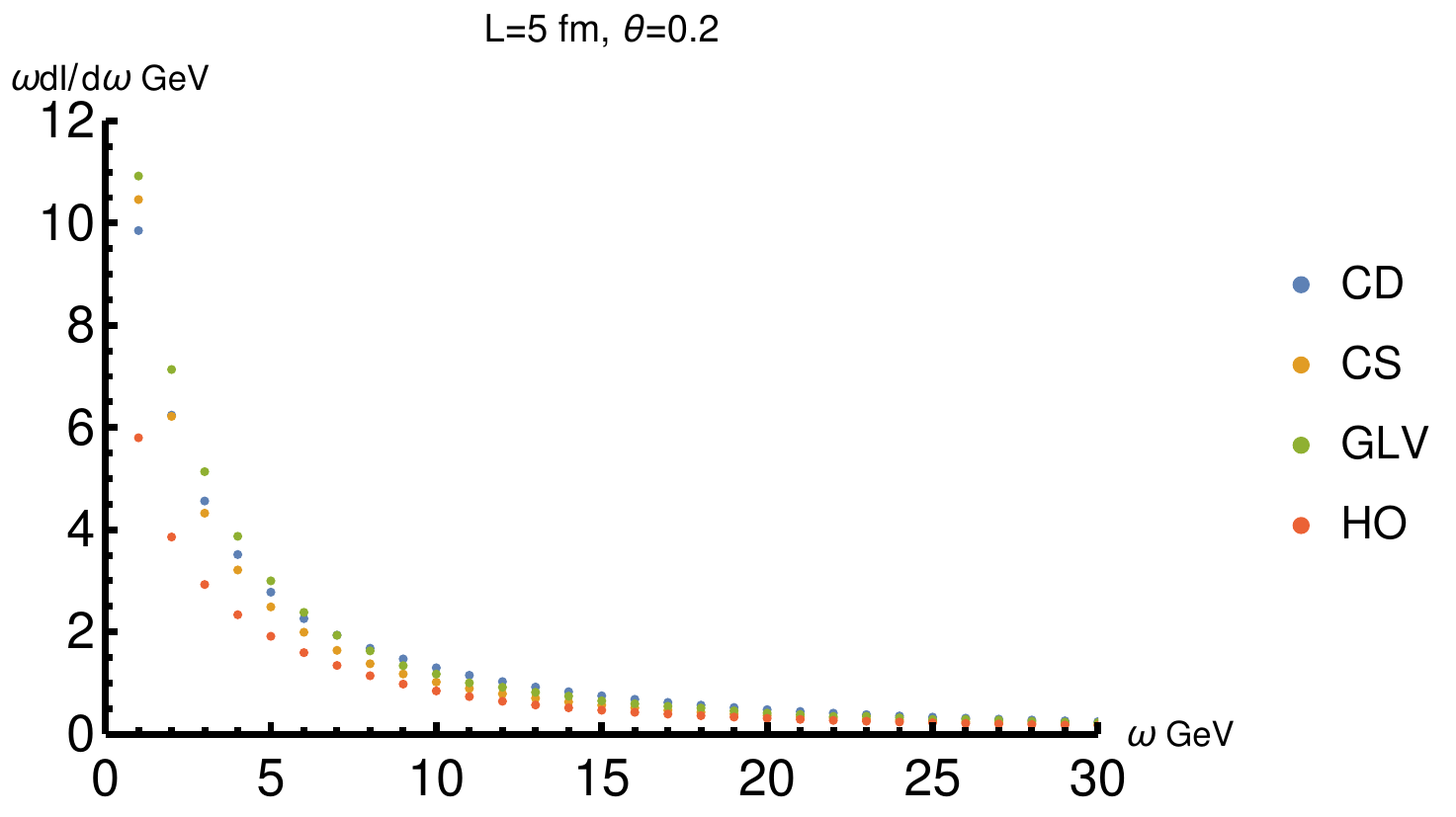}
\caption{\textsl{.   The energy loss  in the leading order in $\alpha_s$ for L=5 fm for different values of $\theta$ as a function of the radiated gluon energy $\omega$, divided by $\alpha_sC_F$ for different
parameterisations of  $Q(\hat q,\omega,\theta)$  for HO and total (HO+Coulomb) contributions: CD, CS  are the total (HO+Coulomb)  energy losses  for typical momenta $Q$ given by Eqs. \ref{m1},\ref{m2} respectively.HO refers to HO approximation 
with $Q$ given by Eq. \ref{m1a}, and GLV refers to N=1 GLV expression with $Q$ independent $\hat q$.}}
\label{2}
\end{figure}

\begin{figure}[t]
\includegraphics[scale=0.53]{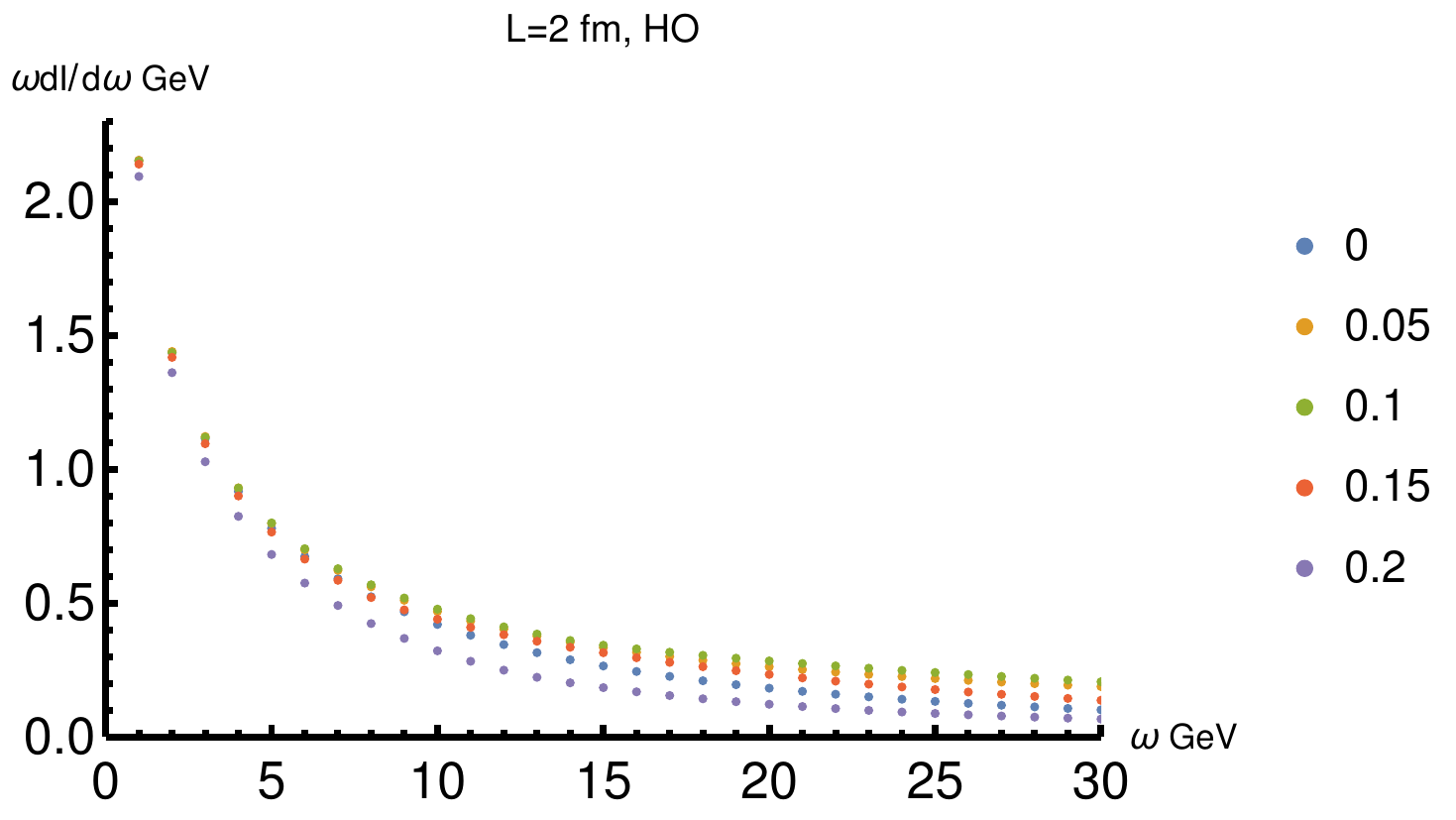}
\includegraphics[scale=0.53]{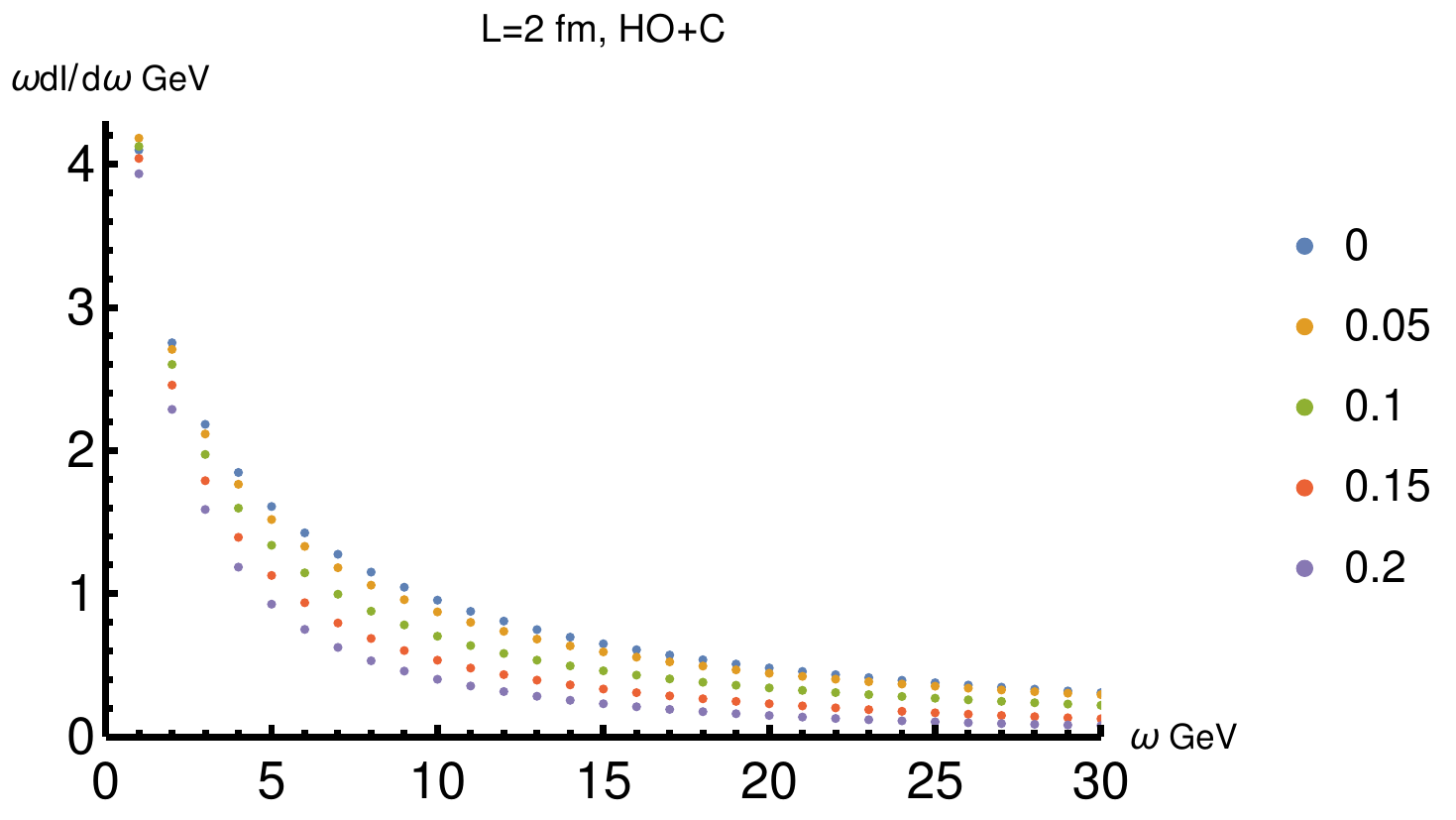}
\caption{\textsl{. The energy loss  in the leading order in $\alpha_s$ for L=2 fm for different values of $\theta$ as a function of the radiated gluon energy $\omega$, divided by $\alpha_sC_F$,  right-total energy loss in Moller theory, left-HO approximation.
We use Eq. \ref{m2} for $Q$.}}
\label{3}
\end{figure}

\begin{figure}[t]
\includegraphics[scale=0.53]{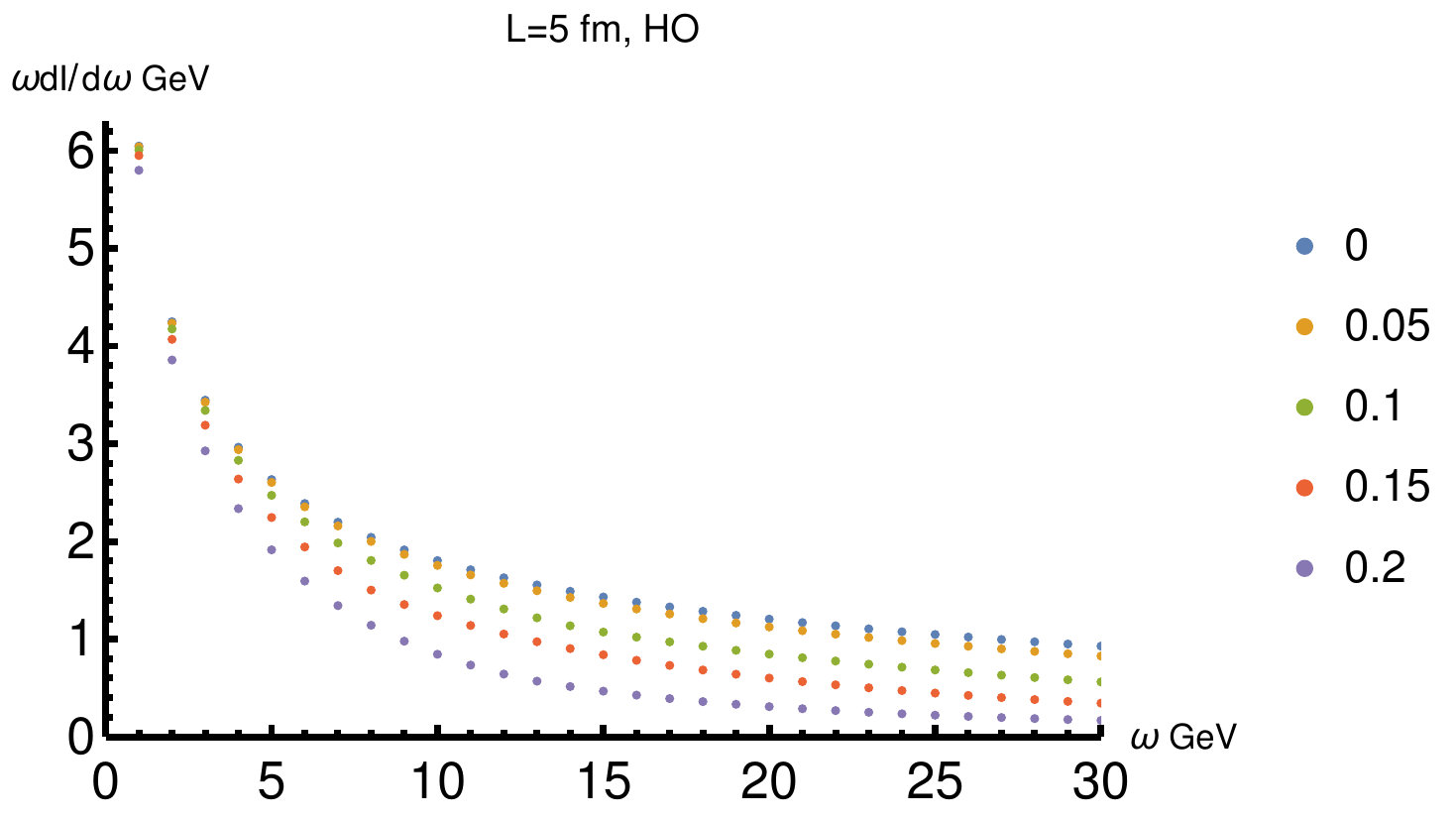}
\includegraphics[scale=0.53]{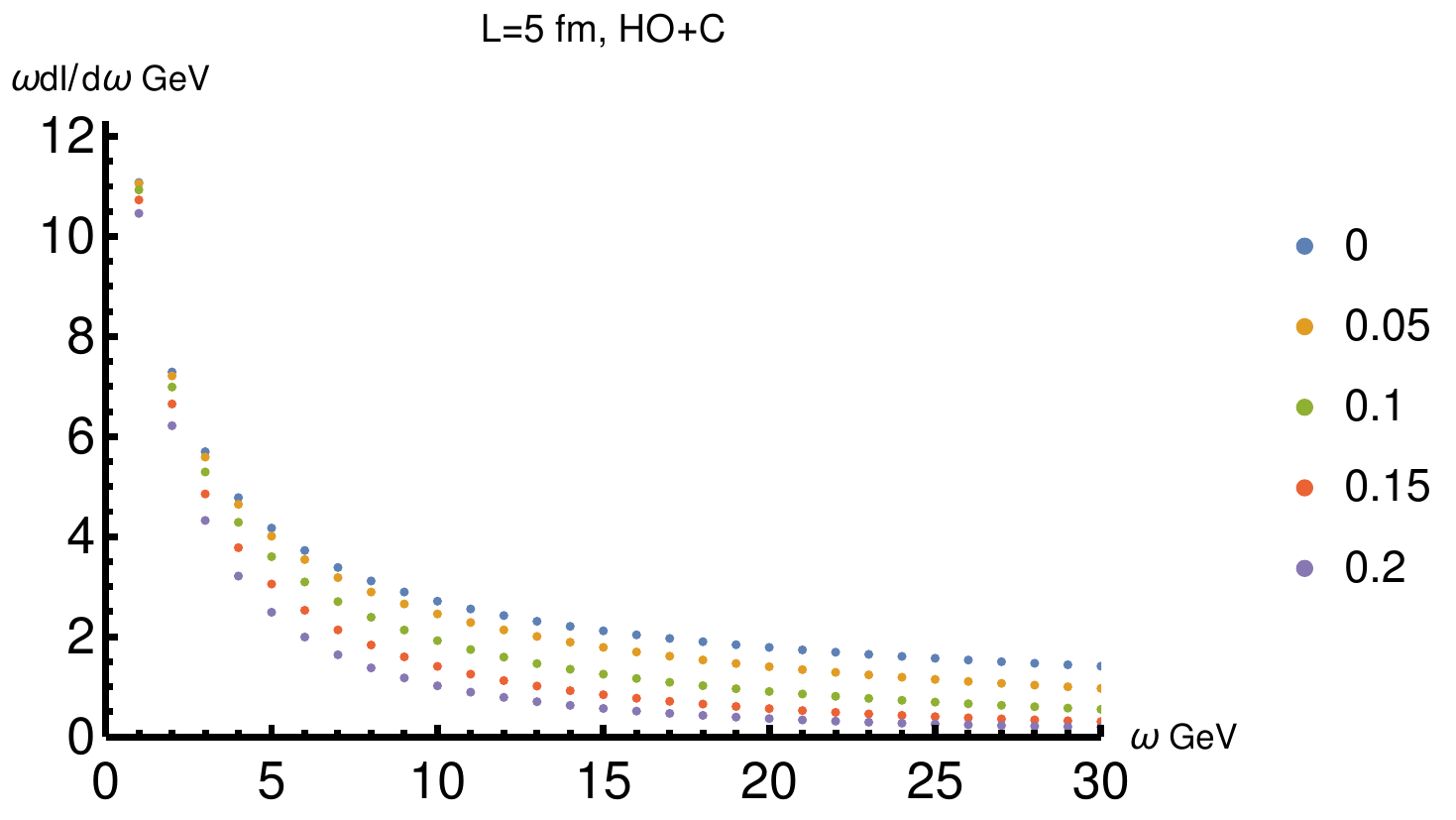}
\caption{\textsl{.   The energy loss  in the leading order in $\alpha_s$ for L=5 fm for different values of $\theta$ as a function of the radiated gluon energy $\omega$, divided by $\alpha_sC_F$  right-total energy loss in Moller theory, left-HO approximation.We use Eq. \ref{m2} for $Q$.}}
\label{4}
\end{figure}
In our calculations we use our final  expression for HO+Coulomb correction to energy loss--Eqs. \ref{cor15}.
For illustrative numerical estimates we take the same parameters as in \cite{mehtar1}:$T=0.4$ GeV, $\alpha_s=0.3$,
leading to $\mu=m_D=0.9$ GeV and $\bar q\sim 0.3$ GeV$^3$.
We carried the numerical calculations for two interpolating formula for typical momentum:
\beq 
Q^2=\sqrt{\omega \hat q_{\rm eff}}U(-\omega+\omega_{DC})+\theta^2\omega^2U(\omega -\omega_{DC})+\mu^2,\label{m1a}
\eeq
\beq
Q^2=\sqrt{\hat q_{\rm ef}\omega+\theta^4\omega^4}+\mu^2,\label{m2a}
\eeq
which differ from Eqs. \ref{m1},\ref{m2}  by adding the regularising momenta $\mu^2$, similar to the regularisation in \cite{mehtar1}.

\par  We present our results for energy loss $\omega dI/d \omega$ for two cases; the medium path width  $L=2$ fm and large path length $L=5$ fm.
We see first that the different choice of the interpolating formula for typical transverse momenta, \ref{m1} or \ref{m2} does not influence the result qualitatively,
although it may induce some difference at small $\omega$ of order 10-15 percent. 
\par For large $\omega$ beyond dead cone frequency $\omega _{DC}$ we see very good agreement with
N=1 GLV approximation, especially for  the intermediate length case. L=2 fm.  We present L=2 fm case in Fig. 1 where we see that the results including coulombic corrections are in very good
agreement with N=1 GLV for large frequencies beyond dead cone. The use of two interpolating formulae for momenta leads to very close results especially they become identical in the dead cone regime.
\par In Fig.2 we depict the similar results for L=5 fm. We see that in this case for small $\omega$ both HO and HO+Coulomb curves lie under the N=1 GLV curve.  We use Eq. \ref{m2} for $Q$.	 curve, with the agreement increasingly good towards large frequencies
 corresponding to the region inside dead cone.
\par In Figs. 3 and 4 we depict for L=2 and 5 fm respectively  the total and HO contributions to the energy loss for different values of $\theta$. We see that for the $\theta$ up to 0,05 the energy loss does not change, but for larger $\theta$ it starts to decrease.
\section{Quenching}
\par Our results for energy loss can be translated to the jet quenching weights along the lines of \cite{BDMPS4},\cite{Wiedemann3}..
As it is known the jet quenching factor describes the energy loss due to the arbitrary number of Poisson distributed gluons.
Indeed, in the previous chapters we calculated the energy loss probability $\omega dI d\omega$ in the first order in $\alpha_s$. 
Then we can calculate the quenching factor
\begin{eqnarray}
Q(E)&=&\exp(-\int^\infty_0(1-\exp(-\frac{{\it R}}{E}\omega)\frac{dI}{d\omega}))\nonumber\\[10pt]
&=&\exp(-\frac{{\it R}}{E}\int^\infty_0\exp(-\frac{{\it R}}{E}N(\omega))\equiv \exp(-S(E))\nonumber\\[10pt]
\label{q1}
\end{eqnarray}
where the multiplicity. 
\beq
N(\omega)=\int^\infty_\omega \omega'\frac{dI}{d\omega'}d\omega'
\eeq
and 
\beq
{\it R}=\frac{d\sigma^{0}}{dp^2_t}
\eeq
is determined from the experimental data, ${\it R}\sim 5$. Here $\sigma^{0}$ is the radiation cross section in the vacuum, outside of the media.
\setlength{\extrarowheight}{0.2cm}
\begin{table}[t]
\begin{center}
\begin{tabular}{cccc}  \hline  \hline 
 & \hspace{0.5cm} $E=25$ GeV \hspace{0.5cm} & \hspace{0.5cm} $E=50$ GeV \hspace{0.5cm} & \hspace{0.5cm} $E=100$ GeV \\  
$L=2$ fm&  S(E) & S(E) & S(E)\\ \hline
Light quark m=0& 1.74 &  1.38 &0.83  \\
Heavy quark $m_b=5$ GeV & 0.75      &  0.79       & 0.77 \\
$L=5$ fm&  S(E) & S(E) & S(E)\\ \hline
Light quark m=0& 4.74 &  3.84 &2.5  \\
Heavy quark $m_b=5$ GeV & 2.55      &  3      & 2.32\\
\hline
\end{tabular}
\caption{\textsl{The estimate for quenching coefficients S(E) for light and heavy quarks, for middle $L=2$ fm and long $L=5$ fm widths. The jet quenching factor $Q(E)=\exp(-S(E))$ Here all S(E) are divided on $\alpha_sC_F$}}
\label{quenching}
\end{center}
\end{table}
The estimated quenching rates have qualitative character, we assumed that for $\omega\le 1$ GeV the $\omega dI/d\omega$ curve linearly goes to zero ar $\omega\rightarrow 0,\omega\le 1$.
.
We see that for energies of order 100 GeV (i.e. $\theta\sim 0.05$ the quenching coefficients are actually the same  for light quark and heavy b-qiuark. and the quenching coefficients 
of heavy quarks depend on energy much weaker than for the light, especially for not large L.
This is in agreement with the results of \cite{Z}.
 The estimates have very qualitative character especially for light quarks, since 
we expect they will be further influenced by phase space restrictions which are known to significantly reduce the energy loss, especially for small $\omega$.
\section{Conclusion}
\par We have calculated the the energy loss of heavy quark propagating through the Quark-Gluon Plasma in the framework of the Moller theory due to the soft gluon emission.
In particular we studied the influence of large Coulomb logarithms on the heavy quark propagation. We have found rather large Coulomb corrections to
LPM effect for small and large energies of radiated gluons. In particular we have seem that Coulombic corrections lead to N=1 GLV expression for energy
loss for frequencies corresponding to the radiation inside the dead cone. Our main expression  that includes both HO approximation and Coulombic logarithms is 
an Eq. \ref{cor15}. We have estimated the resulting quenching weights for heavy quark propagation and see that the energy loss of heavy and light quark is approximately the same up to $\theta=m/E\sim 0.05$, We also see that the difference between heavy and light quenching weights decreases with 
the decrease of the length path of the quark. These results are in agreement with the results of \cite{Z}. For massless quarks our results coincide with those
of \cite{mehtar,mehtar1} .
\begin{acknowledgments}
\noindent
The author thanks  K.~Tywoniuk for useful discussions. The work was supported by Israel Science  Foundation  under  the  grant  2025311.
\end{acknowledgments}

 \end{document}